\documentclass[aps,reprint,pre]{revtex4-1}
\usepackage{graphicx}
\usepackage{subfigure}
\usepackage{epstopdf, epsfig}

\usepackage{amsmath}
\usepackage{dcolumn}
\usepackage{bm}
\usepackage[colorlinks=true, linkcolor=blue]{hyperref}

\begin{document}

\title{Do pulsars rotate clockwise or counterclockwise?}
\author{Renaud Gueroult}
\affiliation{LAPLACE, Universit\'{e} de Toulouse, CNRS, INPT, UPS, 31062 Toulouse, France}
\author{Yuan Shi}
\affiliation{Lawrence Livermore National Laboratory, Livermore, CA 94550, USA}
\author{Jean-Marcel Rax}
\affiliation{Universit\'{e} de Paris XI - Ecole Polytechnique, LOA-ENSTA-CNRS, 91128 Palaiseau, France}
\author{Nathaniel J. Fisch}
\affiliation{Department of Astrophysical Sciences, Princeton University, Princeton, NJ 08540, USA}

\date{\today}

\begin{abstract}
Pulsars are rotating neutron stars which emit lighthouse-like beams. Owing to their unique properties, pulsars are a unique astrophysical tool to test general relativity, inform on matter at extreme densities, and probe galactic magnetic fields. Understanding pulsars physics and emission mechanisms is critical to these applications. Here we uncover that mechanical-optical rotation in the pulsars' magnetosphere affects polarisation in a way which is indiscernible from Faraday rotation in the interstellar medium for typical GHz observations frequency, but which can be distinguished in the sub-GHz band. Besides being essential to correct for possible systematic errors in interstellar magnetic field estimates, our novel interpretation of pulsar polarimetry data offers a unique means to determine whether pulsars rotate clockwise or counterclockwise, providing new constraints on magnetospheric physics and possible emission mechanisms. Combined with the ongoing development of sub-GHz observation capabilities, our finding promises new discoveries, such as the spatial distributions of clockwise rotating or counterclockwise rotating pulsars, which could exhibit potentially interesting, but presently invisible, correlations or features.
\vspace{1cm}
\end{abstract}

\maketitle

Pulsars are strongly magnetized rotating neutron stars. Because of rotation, pulsars emit two intense radiation beams~\cite{Becker2009}. For a distant observer, emission appears as a pulse each time the beam sweeps across his line-of-sight. Owing to their unique properties, pulsars have played, and continue to play, a critical role in the development of astronomy and astrophysics. For instance, pulsars' extreme density makes them one-of-a-kind tools to test both the equation of state of superdense matter~\cite{Sieniawska2018} and the theory of general relativity in the strong field limit~\cite{Kramer2006,Ransom2014,Kramer2016,Shao2018}, while their unparalleled emission stability could allow detecting nanohertz gravitational waves~\cite{Verbiest2016}. Millisecond pulsars also enabled the first detection of an extra-solar planetary system~\cite{Wolszczan1992}.


Pulsars' highly polarised emission and compactness also make them unmatched sources to probe the magnetic fields through Faraday rotation~\cite{Faraday1846}, and pulsars have been instrumental in mapping magnetic field properties in the interstellar medium (ISM) of the Milky Way~\cite{Lyne1968,Han2006,Eatough2013}. These studies generally rely on the assumption that polarisation rotation $\Delta \phi$ results only from the Faraday effect experienced in the non-moving magnetised plasma between the source and the observer. For wave frequency $\omega$ much greater than the plasma frequency $\omega_{pe}$, such as radio-waves in the ISM (see Table~\ref{Tab:Table1}), one can then show that $\Delta \phi^{\textrm{F}}  = RM~\lambda^2$, with $\lambda$ the vacuum wavelength. Information on the magnetic field orientation and strength along the line of sight is then derived from the proportionality coefficient $RM$, called the rotation measure. 

\begin{figure}[htbp]
\caption{Illustration of the different contributions to pulsars' emission polarisation rotation. Polarisation rotation is typically assumed to stem from Faraday rotation between  $\mathcal{Q}$ and $\mathcal{R}$. But wave polarisation also contains information on the magnetosphere properties between points $\mathcal{P}$ and $\mathcal{Q}$, and in particular on the magnetosphere rotation $\Omega$.}
\includegraphics[]{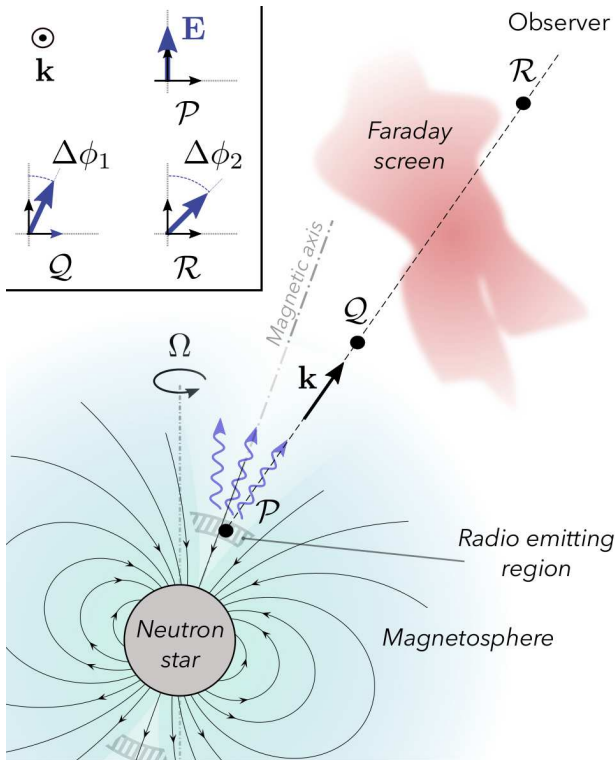}
\label{Fig:Pulsar}
\end{figure}

However, pulsars are surrounded by a magnetosphere. Although pulsar magnetospheric physics, and with it the mechanism responsible for pulsars' emission, remains poorly understood~\cite{Melrose2016,Manchester2004}, it is widely accepted that the magnetosphere is populated by relativistic electron-positron pairs, and that it, or at least its inner region, co-rotates with the neutron star. The analysis of pulsar's signal should hence in principle not only account for propagation in the interstellar medium between the pulsar and the observer (between points $\mathcal{Q}$ and $\mathcal{R}$ in Fig.~\ref{Fig:Pulsar}), but also for propagation in the rotating magnetosphere (between points $\mathcal{P}$ and $\mathcal{Q}$ in Fig.~\ref{Fig:Pulsar}). In particular, pulsar polarimetry ought to consider both the the well known Faraday rotation induced by intervening plasma screens and the possible polarisation rotation in the magnetosphere~\cite{Wang2011}.

\begin{table}[htbp]
\begin{center}
\caption{Typical plasma parameters in the interstellar medium and in pulsars' magnetosphere. Rotation measure $RM$ are typically observed at $\omega\sim1$~GHz. $\omega_{pe}$ and $\omega_{ce}$ are the plasma and electron cyclotron frequency, respectively. }
\begin{tabular}{c | c c c | c c}
 & $B_0$~[T] & $n$~[m$^{-3}$] & $\Omega$~[s$^{-1}$] & $\omega_{pe}/\omega$ & $\omega_{ce}/\omega$ \\
 \hline
Interstellar medium & $10^{-10}$ & $10^6$ & $-$ & $10^{-4}$ & $10^{-8}$ \\
Pulsar magnetosphere & $10^{8}$ & $10^{20}$ & $10$ & $10^{3}$ & $10^{10}$ 
\end{tabular}
\label{Tab:Table1}
\end{center}
\end{table}

\section*{Results}

The effect of the rotating magnetosphere on the polarisation rotation can be both significant and revealing.  First of all, any deduction of the intervening magnetic field between the pulsar and the observer through Faraday rotation will have to be corrected for the additional polarisation rotation.  Second of all, significant new information can be obtained about the pulsar.   For example, absent accounting for the effect of the rotating magnetosphere, the observation of a pulsar from a single distant point will uncover the pulsar rotational frequency, but not whether it is rotating clockwise or counterclockwise. However, as we show here, because the wavelength dependency of the polarisation rotation due to rotating magnetosphere differs from that due to Faraday rotation, these rotation directions can in fact be disambiguated even when observed from a single distant point. This is important because the sign of $\bm{\Omega}\cdot\mathbf{B}$ constrains   possible magnetosphere compositions (electron-positron only or also proton) and particle acceleration mechanisms, and in turn, possible radio emission mechanisms~\cite{Melrose2016}. 

\vspace{0.2cm}\noindent\textbf{Polarisation in rotating gyrotropic media. }To see this, consider for simplicity the propagation of a wave along the axis of an aligned rotator, that is to say when the obliquity $\alpha=0$ in Fig.~\ref{Fig:Pulsar} (i.e., the rotation and magnetic axes are aligned). 
For perfect alignment of the axes ($\alpha=0$ ), the radiation pulsing vanishes, but the key effects are retained. This case is also important since the beam axis tends to align with the rotational axis ($\alpha \rightarrow 0$) as the pulsar ages~\cite{Tauris1998}. In this case, we show (see Methods for the derivation) that polarisation rotation in a rotating gyrotropic medium is the sum of two contributions, with
\begin{equation}
\Delta \phi = \Delta \phi^{\textrm{F}} + \Delta \phi^{\textrm{M}}(\Omega).
\end{equation}
The first term $\Delta \phi^{\textrm{F}}$ is the classical Faraday rotation which occurs in a stationary gyrotropic medium. The second term $\Delta \phi^{\textrm{M}}(\Omega)$ stems from the medium's rotation at frequency $\Omega$, and is referred to as mechanical-optical rotation (MOR)~\cite{Fermi1923,Player1976}. Information on $\Omega$ is therefore imprinted in wave polarisation. This result is expected to hold when the wave propagates along the magnetic field ($\mathbf{k}\parallel\mathbf{B}$), even if the mechanical and magnetic axes are only nearly aligned. Rotation can thus in principle be retrieved from $\Delta \phi^{\textrm{M}}$ in pulsars' pulsating signal.


\vspace{0.2cm}\noindent\textbf{Mechanical-optical rotation in $e-p$ magnetosphere. }In a rotating magnetised plasma, the combined effects of Faraday rotation and MOR makes eliciting the effect of mechanical rotation difficult.  Yet, it happens that Faraday rotation cancels in the particular case of a cold electron-positron ($e-p$) plasma symmetrical in density ($n_{e} = n_{p} = n$). We thus take advantage of this coincidence to shed light onto how polarisation may be affected as a wave propagates in the rotating magnetosphere. 

In an $e-p$ plasma rotating at $\Omega>0$, we show (see Methods) that the left-handed circularly polarised (LCP) wave only propagates above a cut-off frequency $\omega_{lc}$ which depends on $\Omega$ and the $e-p$ plasma density $n$ in the magnetosphere. Above this cut-off frequency, both LCP and RCP waves propagate, and the difference in wave index $\Delta n = n_l-n_r$ introduced by mechanical rotation leads to MOR. Importantly, $\omega_{lc}\sim 10^8$~s$^{-1}$ for plasma parameters typical of pulsar magnetospheres (see Table~\ref{Tab:Table1}), which is on the lower end of the frequency range used for pulsar polarimetry (typically GHz)~\cite{Taylor2009}. This implies that MOR should be present in a large fraction of pulsar polarisation data.

%

\begin{figure}
\caption{Mechanical contribution to polarisation rotation in a rotating symmetrical $e-p$ plasma. $\delta^{M} = \Delta \phi^{M}/l$ is the mechanical-optical rotation (MOR) per unit length. Well above the cut-off $\omega_{lc}$, MOR scales like $\omega^{-2}$, similarly to Faraday rotation. In a limited frequency band above $\omega_{lc}$ ($[\omega/\omega_{lc}-1]\ll1$), a different scaling is found, making it in principle possible to separate mechanical effect from Faraday rotation. $B_0 = 10^8$~T, $n = 10^{20}$~m$^{-3}$ and $\Omega = 10$~s$^{-1}$, $\omega_{lc}\sim 185$~MHz.}
\includegraphics[]{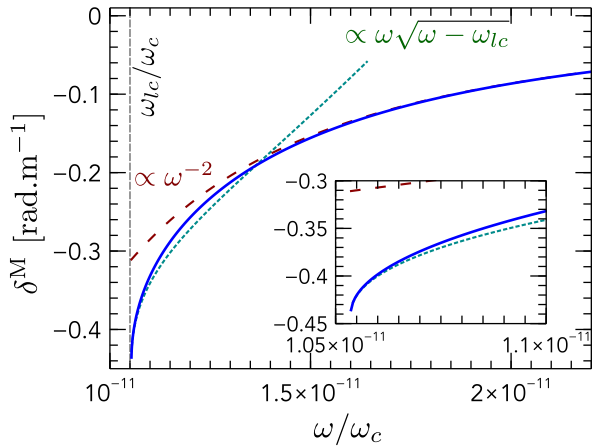}
\label{Fig:Polarization_rotation}
\end{figure}

For frequencies at least a few times $\omega_{lc}$, we see (see Fig.~\ref{Fig:Polarization_rotation}) that $\Delta n(\omega)\propto\omega^{-3}$, and therefore $\Delta \phi^{M}\propto\omega^{-2}$. It follows that MOR in the rotating $e-p$ magnetosphere is a priori indiscernible from Faraday rotation in the intervening interstellar medium since both contributions are proportional to $\lambda^2$. When fitting observations using the relation $\Delta \phi = RM \lambda^2$, the rotation measure $RM$ in this frequency range not only portrays Faraday rotation but also any possible MOR. Attributing $RM$ to the effect of magnetic fields in the interstellar medium alone, as is often done in pulsar polarimetry, thus risks systematic errors. Quantitatively, the magnetic field strength along the line of sight will be respectively over- and under-estimated for negative and positive rotation of the source pulsar since Faraday rotation and MOR are in opposite directions for $\bm{ \Omega}\cdot \mathbf{B}>0$.

\vspace{0.2cm}\noindent\textbf{Low-frequency wavelength-scaling deviation. }However, we uncover here that the peculiar behaviour of MOR near the LCP cut-off may retire this apparent ambiguity. As illustrated in Fig.~\ref{Fig:Polarization_rotation}, mechanical-optical rotation features a different wavelength scaling for $\nu = \omega-\omega_{lc}\ll \omega_{lc}$. In this frequency band, $\Delta n$ increases as $\sqrt{3\nu/\omega_{lc}}$ ($\Delta n<0$ for MOR with $\Omega>0$). For a wave just above the cut-off, polarisation rotation $\Delta \phi_{\mathcal{P}\mathcal{Q}}(\omega)$ between point $\mathcal{P}$ and $\mathcal{Q}$ in Fig.~\ref{Fig:Pulsar} then scales like $\omega\sqrt{\omega-\omega_{lc}}$. In contrast, for the same wave frequency, $\Delta \phi_{\mathcal{Q}\mathcal{R}}(\omega)$ due to Faraday rotation is very well approximated by the classical $\omega^{-2}$ law since $\omega_{lc}$ is over $1000$ times larger than $\omega_{pe}$ in the interstellar medium (see Table~\ref{Tab:Table1}). Combined with the jump in polarisation angle which is predicted to take place at the cut-off frequency, these different frequency scalings offer a conceptual means to separate MOR in the rotating $e-p$ magnetosphere from Faraday rotation in the interstellar medium. MOR could then provide insights into the pulsar magnetosphere dynamics. In particular, a negative deviation from the $\lambda^2$ high-frequency fit above the cut-off will indicate a counter-clockwise rotation of the pulsar magnetosphere, while a positive deviation will indicate clockwise rotation. 

\section*{Discussion}

While this symmetrical $e-p$ plasma magnetosphere model is overly simplified, we show that accounting for relativistic-quantum effects and density asymmetry in the magnetosphere does not qualitatively modify this picture (see Methods). It suggests that while the detailed frequency dependence $\Delta \phi (\omega)$ near the cut-off changes as the magnetosphere model is refined, the existence of a frequency cut-off and the associated non $\lambda^2$ behaviour near this cut-off are robust features of mechanical rotation. 

Our finding that MOR and Faraday rotation show different frequency dependences near the cut-off is particularly relevant in light of the recent observations that polarisation rotation in certain pulsars does not follow the classical $\lambda^2$ law~\cite{Dai2015}, even at higher frequency $f\geq 730~$MHz. If these observed deviations were attributable to MOR, it stands to reason that observations at lower frequency (closer to the predicted $\omega_{lc}$) will reveal a greater fraction of pulsars displaying a non $\lambda^2$ relation. The ongoing development of observation capabilities at radio frequencies below $250~$MHz for gravitational waves detection with pulsar timing arrays~\cite{Verbiest2016} should enable confirming this conjecture by measuring the frequency dependence of pulsars' polarisation closer to the predicted cut-off frequency~\cite{Noutsos2015,Bhat2018}. If successful, this would provide unique means to advance our understanding of the magnetospheric structure and pulsar radio emission mechanism. Finally, if observed deviations were indeed traceable to MOR, polarisation data obtained at shorter wavelengths where $\Delta \phi\propto\lambda^2$ holds true may have to be revisited to correct magnetic field estimates in light of the mechanical rotation contribution.

\section*{Methods}

\vspace{0.2cm}\noindent\textbf{Polarisation rotation in gyrotropic media. }Consider a typical magneto-optic medium described by the susceptibility tensor 
\begin{equation}
\bm{\bar{\chi}}(\bar{\omega}) = \begin{pmatrix}
\bar{\chi}_{\perp} & -j\bar{\chi}_{\times} & 0\\\
j\bar{\chi}_{\times} & \bar{\chi}_{\perp} & 0\\
0 & 0 & \bar{\chi}_{\parallel}
\label{Eq:dielectric_tensor}
\end{pmatrix}.
\end{equation}
The magneto-optical activity translates into right-handed circularly polarised (RCP) and a left-handed circularly polarised (LCP) eigenmodes propagating along $\mathbf{\hat{z}}$, with indices $\bar{n}_r = (1+ \bar{\chi}_{\perp} + \bar{\chi}_{\times})^{1/2}$ and $\bar{n}_l = (1+ \bar{\chi}_{\perp} - \bar{\chi}_{\times})^{1/2}$, respectively. Here left- and right-handed waves are defined from the point of view of the source in the direction of propagation of the wave.

The difference in wave index $n_r$ and $n_l $ of RCP and LCP waves associated with the non zero off-diagonal term $\bar{\chi}_{\times}$ leads to a rotation of the plane of polarisation of a linearly polarised wave. After propagating over a distance $l$, the polarisation has been rotated by 
\begin{equation} 
\Delta \phi(\omega)  = \left[n_l(\omega)-n_r(\omega)\right]\frac{\omega l}{2c}.
\label{Eq:general_polarization_rotation}
\end{equation}
The polarization rotation per unit length, also known as the specific rotary power, is $\delta (\omega) = \Delta \phi(\omega) /l$.

A magnetised plasma can be considered as an anisotropic dielectric. Writing the background magnetic field $\mathbf{B}_0 = B_0\mathbf{\hat{z}}$ and assuming a cold and collisionless plasma, the components of the susceptibility tensor in the plasma rest frame are~\cite{Rax2005} 
\begin{subequations}
\begin{align}
\bar{\chi}_{\perp}(\omega) & = \sum\limits_{\alpha}\frac{{\omega_{p\alpha}}^2}{{\omega_{c\alpha}}^2-\omega^2}\\ 
\bar{\chi}_{\times}(\omega) & = \sum\limits_{\alpha}\epsilon_{\alpha}\frac{\omega_{c\alpha}}{\omega}\frac{{\omega_{p\alpha}}^2}{\omega^2-{\omega_{c\alpha}}^2}\label{Eq:Tensor_component_perp}\\
\bar{\chi}_{\parallel}(\omega) & = -\sum\limits_{\alpha}\frac{{\omega_{p\alpha}}^2}{\omega^2},
\end{align}
\label{Eq:Tensor_components}
\end{subequations}
where $\omega_{c\alpha} = |q_\alpha|B_0/m_\alpha$ and $\omega_{p\alpha} = [n_{\alpha} e^2/(m_{\alpha}\epsilon_0)]^{1/2}$ are the cyclotron frequency and plasma frequency of species $\alpha$, respectively, and $\epsilon_{\alpha} = q_\alpha/|q_\alpha|$. 

Typically, plasma parameters in the Faraday screen in between the pulsar and the observer are such that $\omega_{c\alpha}\ll\omega$ and $\omega_{p\alpha}\ll\omega$ for the GHz wave of radio-telescope measurements (see Table~\ref{Tab:Table1}). In this limit, $1\gg |\bar{\chi}_{\perp}| \gg |\bar{\chi}_{\times}|$, $\bar{\chi}_{\perp}<0$ and $\bar{\chi}_{\times}<0$, so that $n_l(\omega)\geq n_r(\omega)$ and, from Eq.~(\ref{Eq:general_polarization_rotation}), $\Delta \phi>0$. Quantitatively, 
\begin{equation}
n_l(\omega)- n_r(\omega)\sim \frac{\omega_{ce}{\omega_{pe}}^2}{\omega^3},
\end{equation}
which yields the classical scaling $\Delta \phi \propto \lambda^2$.

\vspace{0.2cm}\noindent\textbf{Parallel propagation in rotating gyrotropic media. }Let us now assume that the medium defined by Eq.~(\ref{Eq:dielectric_tensor}) is rotating with $\bm{\Omega} = \Omega \mathbf{\hat{z}}$, and that the dielectric properties in the medium's rest frame are not modified by rotation, \emph{i.~e.} $\bm{\chi'} = \bm{\bar{\chi}}$. Here $p'$ refers to the laboratory frame variable $p$ in the gyrotropic medium's rest frame. In the rotating frame, the constitutive relations write
\begin{subequations}
\begin{align}
\mathbf{B'} & = \mu_0 \mathbf{H'}\\
\mathbf{D'} & = \epsilon_0 \left[\mathbf{I}+\bm{\bar{\chi}} (\omega')\right] \mathbf{E'}. 
\end{align}
\end{subequations}
Using Lorentz transformation from the dielectric rest frame rotating at instantaneous velocity $\mathbf{v}=~^T(-\Omega y, \Omega x,0)$ to laboratory frame (see, \emph{e.~g.}, Ref.~\cite{Landau1975}), 
we get 
the constitutive relations in the lab frame
\begin{subequations}
\begin{align}
\mathbf{B} & = \mu_0 \mathbf{H} -  \frac{\mathbf{v}}{c^2}\times \bm{\bar{\chi}}(\omega')\cdot\mathbf{E}\label{Eq:constitutive_B}\\
\mathbf{D} & = \epsilon_0 \bm{\varepsilon}\cdot\mathbf{E} + \bm{\bar{\chi}}(\omega')\cdot\left( \frac{\mathbf{v}}{c^2}\times \mathbf{H}\right). \label{Eq:constitutive_D}
\end{align}
\label{Eq:constitutive}
\end{subequations}
The second term in Eqs.~(\ref{Eq:constitutive_B}) and (\ref{Eq:constitutive_D}) represent, to first order in $v/c$, the effect of rotation. This set of constitutive relations Eq.~(\ref{Eq:constitutive}) is complemented by Maxwell's equations
\begin{subequations}
\begin{align}
\bm{\nabla}\cdot\mathbf{B} & = 0 \label{Eq:MaxwellB} \\
\bm{\nabla}\cdot\mathbf{D} & = 0 \label{Eq:MaxwellD} \\
\bm{\nabla}\times\mathbf{E} & = -\frac{\partial \mathbf{B}}{\partial t} \label{Eq:MaxwellE} \\
\bm{\nabla}\times\mathbf{H} & = \frac{\partial \mathbf{D}}{\partial t}. \label{Eq:MaxwellH} 
\end{align}
\end{subequations}

Using Eq.~(\ref{Eq:MaxwellH}) into the curl of Eq.~(\ref{Eq:constitutive_B}), and plugging in Eq.~(\ref{Eq:constitutive_D}), one gets
\begin{align}
c\bm{\nabla}\times \mathbf{B} = & \frac{1}{c}\frac{\partial }{\partial t}\left[(\mathbf{I}+\bm{\bar{\chi}}(\omega'))\cdot \mathbf{E}\right] +\frac{\partial }{\partial t}\left[\bm{\bar{\chi}}(\omega')\cdot\left(\bm{\beta}\times\mu_0\mathbf{H}\right)\right] \nonumber\\ & -  \bm{\nabla}\times\left(\bm{\beta}\times \bm{\bar{\chi}}(\omega')\cdot\mathbf{E}\right),
\label{Eq:rotB}
\end{align}
with $\bm{\beta} = \mathbf{v}/c$. To first order in $\beta$, $\mathbf{B}$ can be substituted to $\mu_0\mathbf{H}$ in the second term on the right hand side. Following Player~\cite{Player1976}, we consider the particular case of a wave propagating along the rotation axis, \emph{i.~e.} $\mathbf{k} = k~\mathbf{\hat{z}}$. Eq.~(\ref{Eq:MaxwellB}) and Eq.~(\ref{Eq:MaxwellD}) require respectively that $\mathbf{B}$ and $\mathbf{D}$ are transverse. Eq.~(\ref{Eq:constitutive_B}) and Eq.~(\ref{Eq:constitutive_D})  then imply that $\mathbf{H}$ and $\mathbf{E}$ have longitudinal amplitudes of order $\beta$. To first order in $\beta$, the operator $\bm{\nabla}$ can thus be replaced by $\mathbf{\hat{z}}\partial/\partial z$ when it operates on field quantities~\cite{Player1976}. Under these assumptions, and after some algebra, the last term in Eq.~(\ref{Eq:rotB}) can be rewritten
\begin{equation}
\bm{\nabla}\times\left[\bm{\beta}\times \bm{\bar{\chi}}(\omega')\cdot\mathbf{E}\right] =  \textrm{Q} \left[\bm{\bar{\chi}}(\omega')\cdot \mathbf{E}\right]
\label{Eq:RHS}
\end{equation}
where we have defined the operator
\begin{equation}
\textrm{Q} = \frac{\Omega}{c^2}\left[\textrm{Q}_1\cdot\bm{\nabla}\times +~\textrm{Q}_2\cdot~+ ~\mathbf{\hat{e}_z}\times\right]
\label{Eq:Operator}
\end{equation}
with
\begin{equation}
\textrm{Q}_1 = \begin{pmatrix}
0 & 0 & x\\
0 & 0 & y\\
-x & -y & 0
\end{pmatrix}
\quad
\textrm{and}
\quad
\textrm{Q}_2 = \begin{pmatrix}
0 & 0 & -y\\\
0 & 0 & x\\
y & -x & 0
\end{pmatrix}\frac{\partial}{\partial z}.
\end{equation}
Further derivation shows that the product of the last two terms of the operator $\textrm{A}$ in Eq.~(\ref{Eq:Operator}) with $\bm{\bar{\chi}}(\omega')\cdot \mathbf{E}$ depends only on $\partial E_z/\partial z$, which is negligible to first order in $\beta$ as a result of Eq.~(\ref{Eq:MaxwellD}) and Eq.~(\ref{Eq:constitutive_D}). Using the vector identity Eq.~(\ref{Eq:curlMat}), and noting that $[\bm{\bar{\chi}}(\omega')\cdot\bm{\nabla}]\times\mathbf{E} = \bar{\chi}_{\parallel} \bm{\nabla}\times\mathbf{E}$, Eq.~(\ref{Eq:RHS}) then writes to first order in $\beta$
\begin{equation}
\bm{\nabla}\times\left[\bm{\beta}\times \bm{\bar{\chi}}(\omega')\cdot\mathbf{E}\right] = \frac{\Omega}{c} \textrm{Q}_1 \bm{\bar{\chi^{\dagger}}} \cdot\left(\bm{\nabla}\times \mathbf{E}\right)
\label{Eq:RHS_simple}
\end{equation}
with 
\begin{equation}
\bm{\bar{\chi}^{\dagger}} = \begin{pmatrix}
\bar{\chi}_{\perp} & -j\bar{\chi}_{\times} & 0\\\
j\bar{\chi}_{\times} & \bar{\chi}_{\perp} & 0\\
0 & 0 & 2\bar{\chi}_{\perp}-\bar{\chi}_{\parallel}
\end{pmatrix}.
\end{equation}
Using Eq.~(\ref{Eq:MaxwellE}) in Eq.~(\ref{Eq:RHS_simple}), plugging it into Eq.~(\ref{Eq:rotB}), and taking the curl, we get
\begin{align}
c\bm{\nabla}\times\bm{\nabla}\times \mathbf{B} = & \frac{1}{c}\frac{\partial }{\partial t}\left(\bm{\nabla}\times\left[\mathbf{I}+\bm{\bar{\chi}}(\omega')\right]\cdot \mathbf{E} \right)\nonumber\\ 
 & + \frac{1}{c}\frac{\partial }{\partial t}\left[\bm{\nabla}\times\bm{\bar{\chi}}(\omega')\cdot\left(\bm{\beta}\times\mathbf{B}\right)\right]\nonumber\\
 &  +  \frac{\Omega}{c} \bm{\nabla}\times\textrm{Q}_1 \bm{\bar{\chi^{\dagger}}}(\omega') \frac{\partial \mathbf{B}}{\partial t}.
\label{Eq:rotrotB}
\end{align}
Using once more the vector identity Eq.~(\ref{Eq:curlMat}), and $([\mathbf{I}+\bm{\bar{\chi}}(\omega')]\cdot\bm{\nabla})\times\mathbf{E} = (1+\bar{\chi}_{\parallel}) \bm{\nabla}\times\mathbf{E}$, the first term in the bracket on the right hand side of Eq.~(\ref{Eq:rotrotB}) reads 
\begin{equation}
\bm{\nabla}\times\left[\mathbf{I}+\bm{\bar{\chi}}(\omega')\right]\cdot \mathbf{E} = \left[\mathbf{I}+\bm{\bar{\chi}^{\dagger}}(\omega')\right]\cdot\bm{\nabla}\times\mathbf{E}.
\label{Eq:d2B}
\end{equation}
Finally, plugging Eq.~(\ref{Eq:MaxwellE}) in Eq.~(\ref{Eq:d2B}), a wave equation for $\mathbf{B}$ is obtained,
\begin{align}
\bm{\nabla}\times\bm{\nabla}\times \mathbf{B} = & -\frac{1}{c^2}\left[\mathbf{I}+\bm{\bar{\chi}^{\dagger}}(\omega')\right]\cdot\frac{\partial^2 \mathbf{B}}{\partial t^2}\nonumber\\
& + \frac{1}{c}\frac{\partial}{\partial t}\bm{\nabla}\times\bm{\bar{\chi}}(\omega')\cdot\left(\bm{\beta}\times\mathbf{B}\right)\nonumber\\
& +  \frac{\Omega}{c^2} \bm{\nabla}\times\textrm{Q}_1 \bm{\bar{\chi^{\dagger}}}(\omega') \frac{\partial \mathbf{B}}{\partial t}.
\label{Eq:waveB}
\end{align} 
Writing $\mathbf{B} =~^\textrm{T}(B_x,B_y,0)\exp[j(k z -\omega t)]$ and introducing the wave index $n = kc/\omega$, Eq.~(\ref{Eq:waveB}) leads to 
\begin{equation}
\begin{pmatrix}
1+\mathbf{\chi_{\perp}}-n^2 & -j\mathbf{\chi_{\times}}\\
j\mathbf{\chi_{\times}} & 1+\mathbf{\chi_{\perp}}-n^2
\end{pmatrix}
\begin{pmatrix}
B_x\\
B_y
\end{pmatrix}
=
\begin{pmatrix}
0\\
0
\end{pmatrix}
\label{Eq:dispersion_B}
\end{equation}
with
\begin{subequations}
\begin{align}
\mathbf{\chi_{\perp}} & = \bar{\chi}_{\perp} -\frac{\Omega}{\omega}\bar{\chi}_{\times} \\
\mathbf{\chi_{\times}} & = \bar{\chi}_{\times}-\frac{\Omega}{\omega}\left(\bar{\chi}_{\parallel}+\bar{\chi}_{\perp}\right). \label{Eq:chi_cross_rot}
\end{align}
\end{subequations}
In deriving Eq.~(\ref{Eq:dispersion_B}), terms in $\partial\bm{\beta}/\partial t$ and $\partial^2\bm{\beta}/\partial t^2$ have been neglected since they are respectively of order $\beta^2$ and $\beta^3$. 

\vspace{0.2cm}\noindent\textbf{Mechanical contribution to polarisation rotation. }From Eq.~(\ref{Eq:dispersion_B}), we see that the wave indexes for RCP ($B_y = jB_x$) and LCP ($B_y = -jB_x$) waves are modified by rotation and now write
\begin{subequations}
\begin{align}
{n_r}^2(\omega) & = 1+\chi_{\perp}(\omega')+\chi_{\times}(\omega')\nonumber\\
 & = 1+ \bar{\chi}_{\perp}(\omega') + \bar{\chi}_{\times}(\omega')\nonumber\\
 & \qquad -\frac{\Omega}{\omega}\left[\bar{\chi}_{\times}(\omega') + \bar{\chi}_{\parallel}(\omega')+\bar{\chi}_{\perp}(\omega')\right],
\label{Eq:index_general_r}
\end{align} 
and
\begin{align}
{n_l}^2(\omega) & = 1+\chi_{\perp}(\omega')-\chi_{\times}(\omega')\nonumber\\
 & = 1+ \bar{\chi}_{\perp}(\omega') - \bar{\chi}_{\times}(\omega')\nonumber\\
 & \qquad -\frac{\Omega}{\omega}\left[\bar{\chi}_{\times}(\omega') - \bar{\chi}_{\parallel}(\omega')-\bar{\chi}_{\perp}(\omega')\right].
\label{Eq:index_general_l}
\end{align}
\label{Eq:index_general}
\end{subequations}
Owing to Doppler shift, $\omega ' = \omega-\Omega$ for the RCP, and $\omega' = \omega+\Omega$ for the LCP.

Just like polarisation rotation in a stationary gyrotropic medium arose from $\bar{\chi}_{\times}\neq 0$, Eq.~(\ref{Eq:index_general}) shows that polarisation rotation in a rotating gyrotropic medium stems from $\chi_{\times}\neq 0$. However Eq.~(\ref{Eq:chi_cross_rot}) indicates that polarisation rotation can now stem either from anisotropy of the medium ($\bar{\chi}_{\times}\neq 0$) or from mechanical rotation ($\Omega\neq 0$), or a combination of the two effects. 

In the limit of an isotropic dielectric, $\bar{\chi}_{\perp}=\bar{\chi}_{\parallel} = \epsilon_r-1$, with $\epsilon_r$ the dielectric relative permittivity, and $\bar{\chi}_{\times} = 0$. Polarisation rotation hence results only from mechanical rotation. Assuming slow rotation ($\Omega\ll\omega$), Eqs.~(\ref{Eq:index_general}) rewrite
\begin{equation}
{n_{l/r}}(\omega) \sim \sqrt{\epsilon_r(\omega')}\pm \left[\sqrt{\epsilon_r(\omega')}-\frac{1}{\sqrt{\epsilon_r(\omega')}}\right]\frac{\Omega}{\omega}.
\end{equation}
Taylor expanding the refractive index difference $\Delta n = n_l-n_r$, one recovers from Eq.~(\ref{Eq:general_polarization_rotation}) the result 
\begin{equation} 
\Delta \phi  = \frac{\Delta n\omega l}{c} = \left(n_g-n^{-1}\right)\frac{\Omega l}{c}
\end{equation} 
first obtained by Player~\cite{Player1976} and later generalised by Goette~\cite{Goette2007} to account for wave optical angular momentum. Here $n_g = n + \omega dn/d\omega$ is the group index and $n^2=\epsilon_r$.

\vspace{0.2cm}\noindent\textbf{Polarisation rotation in a rotating symmetrical electron-positron magnetosphere. }For a symmetrical and cold electron-positron ($e-p$) plasma, $n = n_e = n_p$, $\epsilon_{e} = -\epsilon_{p} = 1$ and $m = m_p = m_e$. The non-diagonal term $\bar{\chi}_{\times}$ of the susceptibility tensor in Eq.~(\ref{Eq:Tensor_component_perp}) hence also cancels. Electrons and positrons interact symmetrically with RCP and LCP waves, respectively, and no polarisation rotation is found in the absence of rotation (no Faraday rotation). 
Polarisation rotation is in this case a purely mechanical effect, as it is the case for an isotropic dielectric~\cite{Player1976}. 

For the typical pulsar magnetosphere parameters given in Table~\ref{Tab:Table1} 
and GHz radio waves typically used by radio-telescopes, the ordering $\omega_c\gg\omega_p\gg\omega\gg\Omega$ holds. In these conditions, $|\bar{\chi}_{\parallel}|\gg 1 \gg\bar{\chi}_{\perp}$. Since $\bar{\chi}_{\parallel}<0$, Eq.~(\ref{Eq:index_general_l}) indicates that there is a cut-off frequency 
\begin{equation}
\omega_{lc} \sim \left(2{\omega_p}^2 \Omega\right)^{1/3}
\end{equation}
below which the LCP does not propagate assuming $\Omega>0$. Note that reversing the pulsar sense of rotation ($\Omega<0$) simply changes the LCP cut-off into a RCP cut-off at the same frequency. For the parameters given here, $\omega_{lc}\sim 10^8$~s$^{-1}$, which is on the lower end of radio-telescope observations. For $\Omega>0$, $n_r(\omega)\geq n_l(\omega)$ above $\omega_{lc}$, and Eq.~(\ref{Eq:general_polarization_rotation}) shows that $\Delta \phi^M<0$. Conversely, $\Delta \phi^M>0$ for $\Omega<0$. Depending on the pulsar sense of rotation, mechanical-optical rotation in the rotating $e-p$ magnetosphere can hence add to or subtract from polarisation rotation associated with the magneto-optical effect in the low-density Faraday screen between the pulsar and the observer. Since the situation is symmetrical, we consider the case $\Omega>0$ in the rest of this section.

Far above the cut-off frequency, that is to say for $\omega_{lc}\ll\omega\ll \left(2{\omega_c}^2 \Omega\right)^{1/3}$, $1\gg |\bar{\chi}_{\parallel}|\Omega/\omega\gg\bar{\chi}_{\perp}$ and
\begin{equation}
n_l(\omega)-n_r(\omega) \sim \frac{\Omega}{\omega}\bar{\chi}_{\parallel} \sim -2\frac{{\omega_p}^2\Omega}{\omega^3}.
\end{equation}
From Eq.~(\ref{Eq:general_polarization_rotation}), polarisation rotation $\Delta \phi$ is hence proportional to $\omega^{-2}$, similarly to Faraday rotation in a stationary magnetised plasma for wave frequencies much larger than the plasma frequency $\omega_{pe}$. 

Interestingly, different behaviour is found near the cut-off. Taylor expanding the left and right wave indexes, one finds, to lowest order in $\nu = \omega-\omega_{lc}$,
\begin{equation}
n_l(\omega)-n_r(\omega) = -\sqrt{2} + \sqrt{3\frac{\nu}{\omega_{lc}}} + \mathcal{O}\left(\frac{\nu}{\omega_{lc}}\right).
\end{equation}
In this frequency band, polarisation rotation $\Delta\phi$ hence scales like $\omega\sqrt{\omega-\omega_{lc}}$. 

Finally, a jump in polarisation angle $\Delta \phi_{lc}$ should be observed when passing $\omega=\omega_{lc}$ for $\Omega>0$. Below the cut-off frequency, only the RCP wave propagates, and the change in polarisation angle after propagating a distance $l$ will hence be $\Delta \phi = -n_r \omega l/c$. The polarisation angle should hence jump by \begin{equation}
\Delta \phi_{lc} = \frac{n_r(\omega_{lc})\omega_{lc}l}{2c}\sim\left(\frac{{\omega_p}^2 \Omega}{\sqrt{2}}\right)^{1/3}\frac{l}{c}
\end{equation}
when $\omega$ increases past $\omega_{lc}$.

\vspace{0.2cm}\noindent\textbf{Effects of magnetosphere model refinement.} While the symmetrical $e-p$ plasma model used so far conveniently highlights the role of mechanical rotation, it fails to account for two features which are typical of pulsar's magnetosphere. 

\begin{figure}[htbp]
\caption{Comparison of polarisation rotation per unit length predictions $\delta = \Delta \phi/l$ obtained for different $e-p$ magnetosphere models. While QED corrections and $e-p$ density asymmetry do affect polarisation rotation near the cut-off, all three cases are found to deviate from the $\omega^{-2}$ scaling. The symmetrical case is the baseline computed for $B_0 = 10^8$~T, $n = 10^{20}$~m$^{-3}$ and $\Omega = 10$~s$^{-1}$. The non-symmetrical is computed for $n_p = n$ and $n_e = (1-f)/fn$ with $f=0.49$.}
\begin{center}
\includegraphics[]{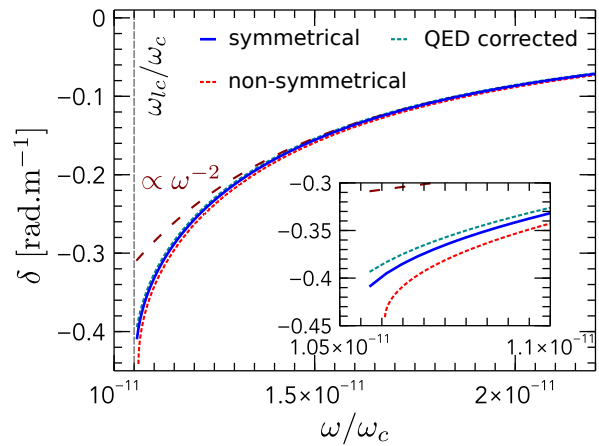}
\end{center}
\label{Fig:Polarization_rotation_compare}
\end{figure}

First, magnetospheres are generally assumed to have non-zero space charge, so that $n_e\neq n_p$. The density asymmetry leads to non-zero non-diagonal susceptibility $\bar{\chi}_{\times}$. This makes polarisation rotation more complicated with now both Faraday rotation and MOR taking place in the magnetosphere. If the charge density is equal to the Goldreich-Julian value $N_{GJ}$~\cite{Goldreich1969}, the relation $\eta(1-2f)=1$ holds true with $f=n_p/(n_e+n_p)$ the positron fraction and $\eta=(n_e+n_p)/N_{GJ}\geq1$. The multiplicity factor $\eta$ is generally assumed to be large ($10^2-10^5$), so that $f$ is close to $0.5$. To illustrate the effect of density asymmetry, we choose $n_p = n$ and $n_e = (1-f)/fn$ with $f=0.49$. This corresponds to a space charge larger than the Goldreich-Julian value for the pulsar parameters given in Table~\ref{Tab:Table1} and used in the symmetrical model for which $\eta = 285$ so that $f\sim 0.498$. Yet, as illustrated in Fig.~\ref{Fig:Polarization_rotation_compare}, we see that there still exists a cut-off for the LCP wave ($\Omega>0$ here), and that the deviation of polarisation rotation near the cut-off from the $\lambda^2$ scaling persists. The observed upshift in cut-off frequency stems from the increase in $\omega_{pe}$.

Second, the relativistic-quantum effects associated with the extremely strong magnetic fields found in pulsars should also be considered~\cite{Shi2016}. The plasma susceptibility tensor Eq.~(\ref{Eq:Tensor_components}) is then replaced by its QED form~\cite{Shi2018}
\begin{subequations}
\begin{align}
\label{eq:QEDchi_perp}
\bar{\chi}_\perp =&-\sum_{\alpha}\frac{m_\alpha{\omega_{p\alpha}}^2}{m_{\alpha 0}\omega^2}\frac{\mathcal{N}_{\alpha}(\omega,n)}{\mathcal{D}_{\alpha}(\omega,n)},\\
\label{eq:QEDchi_cross}
\bar{\chi}_\times =&-\sum_{\alpha}\epsilon_\alpha\frac{\omega_{c\alpha}{\omega_{p\alpha}}^2}{\omega^3}
\frac{4 m_{\alpha}^2}{\mathcal{D}_{\alpha}(\omega,n)},\\
\label{eq:QEDchi_para}
\bar{\chi}_\parallel =&-\sum_{\alpha}\frac{m_\alpha{\omega_{p\alpha}}^2}{m_{\alpha 0}\omega^2}
\frac{\kappa^2\omega^2(1-n^2)-4{m_{\alpha 0}}^2}{\kappa^2\omega^2(1-n^2)^2 -4 {m_{\alpha 0}}^2},
\end{align}
\end{subequations}
with 
\begin{align}
&\mathcal{N}_{\alpha}(\omega,n) = \kappa^2\omega^2(1-n^2)^2-2\kappa(1-n^2)m_\alpha\omega_{c\alpha}-4 {m_{\alpha 0}}^2,
\nonumber\nonumber\\
&\mathcal{D}_{\alpha}(\omega,n) = [\kappa\omega(1-n^2)-2 m_\alpha\omega_{c\alpha}/\omega]^2 -4 {m_{\alpha 0}}^2.\nonumber
\end{align}
Here $\kappa= \hbar/c^2$ and $m_{\alpha 0}=\sqrt{m_\alpha^2+eB_0\hbar/c^2}$ the shifted ground-state mass of the charged particle. Compared to the classical model, the components of the susceptibility tensor now depend on the wave vector $k$, but implicit expressions can be found
for the wave refractive indexes $n_r$ and $n_l$. The numerical solution for our default pulsar parameters is shown in Fig.~\ref{Fig:Polarization_rotation_compare}. This result shows that the deviation from the $\lambda^2$ scaling near the cut-off frequency persists even when QED effects are taken into consideration.

\vspace{0.2cm}\noindent\textbf{Vector identity.} For a function $\mathbf{f}:\textrm{I\!R}^3\rightarrow\textrm{I\!R}^3$ and a constant matrix $\textrm{M}$ with elements $m_{ij}$, $1\leq i\leq3$, the relation
\begin{equation}
\bm{\nabla}\times(\textrm{M}~ \mathbf{f}) =  \textrm{N}\cdot (\bm{\nabla}\times\mathbf{f}) - (\textrm{M}\cdot\bm{\nabla})\times\mathbf{f},
\label{Eq:curlMat}
\end{equation}
holds with
\begin{align}
\textrm{N} & = \begin{pmatrix} m_{22}+m_{33} & -m_{21} & -m_{31}\\ -m_{12} & m_{11}+m_{33} & -m_{32}\\ -m_{13} & -m_{23} & m_{11}+m_{22}\end{pmatrix}\nonumber \\
 & = tr(\textrm{M}) \textrm{I}-\textrm{M}^{\textrm{T}}.\nonumber
\end{align}



%



\vspace{0.2cm}
\noindent\textbf{Acknowledgements } Y. Shi's work was performed under the auspices of the U.S. Department of Energy at Lawrence Livermore National Laboratory under Contract DE-AC52-07NA27344 and was supported by the Lawrence Fellowship through LLNL-LDRD Program under Project No. 19-ERD-038. NJF was supported, in part, by NNSA Grant No. DE-NA0002948.
\vspace{0.2cm}\\
\noindent\textbf{Competing Interests }The authors declare that they have no competing financial interests.\vspace{0.2cm}\\
\noindent\textbf{Correspondence }Correspondence and requests for materials should be addressed to Renaud Gueroult.~(email: renaud.gueroult@laplace.univ-tlse.fr).


\end{document}